\documentclass[11pt]{article}
\usepackage{epsfig}  
     
\begin{document}    

\begin{titlepage}   

\begin{center}
\textbf{\Large A class of spatio-temporal and causal stochastic
       processes, with application to multiscaling and multifractality}
\\[18pt]

J\"urgen Schmiegel, Ole E.\ Barndorff-Nielsen\\
\textit{Thiele Centre for Applied Mathematics in Natural Science,}\\
\textit{Aarhus University,}  \\
\textit{DK--8000 Aarhus, Denmark} \\[12pt]
and\\[12pt]
Hans C.\ Eggers\\
\textit{Department of Physics,} \\
\textit{University of Stellenbosch,} \\
\textit{ZA--7600 Stellenbosch, South Africa}\\[18pt]
\end{center}

\begin{abstract}    
  We present a general class of spatio-temporal stochastic processes
  describing the causal evolution of a positive-valued field in space
  and time. The field construction is based on independently scattered
  random measures of L\'evy type whose weighted amplitudes are
  integrated within a causality cone.  General $n$-point correlations
  are derived in closed form. As a special case of the general
  framework, we consider a causal multiscaling process in space and
  time in more detail. The latter is derived from, and completely
  specified by, power-law two-point correlations, and gives rise to
  scaling behaviour of both purely temporal and spatial higher-order
  correlations. We further establish the connection to classical
  multifractality and prove the multifractal nature of the
  coarse-grained field amplitude.
\end{abstract}


\end{titlepage}

\section*{Nonspecialist summary}     

\begin{quote}
  \noindent
  \textbf{ Throwing many dice many times at different points in space
    is an example of an \emph{uncorrelated random process}, because
    the number of points on a particular die is independent of those
    on other dice around it, and because the dice do not have any
    memory.  Such uncorrelated random processes may seem unsuitable
    for describing correlated phenomena in nature. L\'evy-based
    modelling, however, accomplishes exactly that. It does so by
    making use of overlapping sums of dice as follows: Suppose we have
    three dice labeled $A$, $B$ and $C$ whose outcomes are
    uncorrelated. The two variables $X = A+B$ and $Y = B+C$ will
    nevertheless be correlated since $X$ and $Y$ share the outcome of
    die $B$.  This simple example can be generalised to construct
    correlated processes in spacetime. Let, for example, the energy
    $\varepsilon$ at a point $(x,t)$ be determined by the sum of
    outcomes of all dice occurring within its \textit{ambit set}, a
    kind of causality cone similar to Einstein's familiar light cone
    (see Figure 1). As shown in Figure 2, two energies at different
    points will then be correlated if their respective ambit sets
    overlap, because they will share a common ancestry to some extent.
    The freedom to choose both the kind of randomness and the form and
    size of the ambit set permits this approach to mimic different
    types of correlated behaviour.  This article deals with the
    particular class of phenomena called multifractal and
    multiscaling, which includes turbulence, data traffic flows, cloud
    distributions, rain fields and tumour profiles, to name but a few.
    All of these show a power-law-like behaviour in their correlations
    which are easily incorporated into the L\'evy-based modelling
    scheme by taking \textit{products} of random variables rather than
    their sums. We show how to construct both the kind of randomness
    (the L\'evy basis) and suitable ambit sets for this class, taking
    as input the measured correlations, and calculate analytically
    correlations for overlaps of various kinds (see Figures 2 and 3).
    While this paper concerns itself chiefly with multifractals, the
    L\'evy scheme as such is much more general and can be appled to
    many other correlated random processes.  }
\end{quote}

\section{Introduction}     

Multifractality \cite{FED88} has in the last decade become one of a
number of well-established approaches to the analysis of time series
and spatial patterns, whether nonlinear, random, deterministic, or
chaotic. It serves, for example, to characterize the intermittent
fluctuations observed in fully developed turbulent flows
\cite{MEN91,FRI95} and in data traffic flows of communication networks
\cite{PAR00}. Spatial patterns of cloud distributions and rain fields
\cite{SCHER92,SCHER85} reveal multifractal properties, as do
super-rough tumor profiles \cite{BRU98}. While still controversial,
multiscaling has also been applied to financial time series of
exchange rates and stock indices \cite{MUZ00,CAL02,BAR01}. Many other
examples may be found in the literature.

Multifractality should not, however, be seen as a mere tool for
analysis and characterization: it has also found its way into
theoretical modeling. Maybe the simplest construction of a
multifractal field is achieved with random multiplicative cascade
processes \cite{MEN91} which introduce a hierarchy of scales and
multiplicatively redistribute a flux density from large to small
scales.

Various generalizations of such purely spatial and discrete cascade
processes towards continuous cascade processes in time and/or space,
formulated in terms of integrals over an uncorrelated noise field,
have been undertaken recently.  A purely temporal and causal
generalization to a continuous cascade process is, for example,
discussed by Schmitt \cite{SCHMITT03}, who introduces a log-normal
field, itself defined as an integral of a weighted and uncorrelated
noise field over an associated time-dependent interval. By judicious
choice of integration interval and weight function, the resulting
process is stationary and exhibits approximate scaling behaviour of
two-point correlations.

Muzy and Bacry \cite {MUZ02} discuss a similar approach, constructing
a purely temporal multifractal measure with the help of a limiting
process. Since, however, the field amplitude depends on times later
than the observation time $t$, the model does not obey causality.

The proper description and modeling of spatio-temporal multifractal
physical processes clearly calls for a model generalization that is
causal, explicitly depends on space and time and does so in a
continuous framework. A first step in this direction was achieved in
\cite{SCH03a}, where a continuous and causal spatio-temporal process
was constructed in analogy to a discrete cascade process. Analytical
forms for two- and three-point correlations for the case of a stable
noise field were successfully compared to the corresponding
experimental statistics in fully developed turbulent shear flow.

The aim of the work presented here is to provide a general framework
for the construction of spatio-temporal processes that permits a
unified description of the above-mentioned models
\cite{SCHMITT03,MUZ02,SCH03a} while transcending them all.  The basic
notion in this framework is that of independently scattered random
measures of L\'evy type. The appealing mathematics behind these
measures, as described in \cite{SCH03b} (with emphasis on
spatio-temporal modeling), provide a characterisation of arbitrary
$n$-point correlations independent of the choice of a concrete
realisation of the model. This opens up the possibility of designing
spatio-temporal processes almost to order, i.e.\ satisfying prescribed
correlations.

As an application, we present the construction of a multiscaling and
causal spatio-temporal process that is based on and derived from
scaling two-point correlations. In contrast to \cite{SCHMITT03} and
\cite{SCH03a}, where the specification of the probability density of
the noise-field must be included from the very beginning, we can
construct the process without fixing the marginal distribution of the
field-amplitude. This opens up the possibility of tailoring the
marginal distribution of the process to the phenomenology of a given
application. In particular, the special case of a stable law coincides
with \cite{SCH03a}.

While we will concentrate on multifractal examples in most of this
paper, it should be noted that this framework is not restricted to
multiscaling (defined as scaling of correlation functions) or
multifractal processes (defined as scaling of the coarse-grained
process).

The paper is structured as follows.  In Section \ref{sec:genmod}, we
discuss the general framework for spatio-temporal modeling and derive
an explicit expression for $n$-point correlation functions for the
general set-up. Based on this result, we turn to the application in
the context of causal and multiscaling spatio-temporal processes in
Section \ref{sec:multiscal}, where we show in detail the multiscaling
properties of temporal and spatial $n$-point correlations of arbitrary
order and establish a relation between spatio-temporal multiscaling
and spatio-temporal multifractality.  Section \ref{sec:concl}
concludes the paper with a summary and a brief outlook.

\section{General model approach}     
\label{sec:genmod}

The aim of this Section is to define the general framework and to
provide useful mathematics for the construction of a class of causal
spatio-temporal processes that are based on the integration of an
independently scattered random measure of L\'evy type. The integral
constituting a given observable extends over a finite domain in
space-time, called the \textit{ambit set} $S$.  This approach includes
the special case of a continuous cascade process in space and/or time
as an example. In particular, we recover the temporal cascade
processes discussed in \cite{SCHMITT03} and \cite{MUZ02}, as well as
the spatio-temporal cascade process derived in \cite{SCH03a}. In this
paper, we restrict ourselves to causal processes in $1+1$ dimensions
only, referring the reader to \cite{SCH03b} and \cite{SCH02} for the
general case of $(n{+}1)$-dimensional processes and its various
applications and properties.

The basic notion is that of an \textit{independently scattered random
  measure} (\rm{i.s.r.m}) on continous space-time, $\mathbf{R}{\times}
\mathbf{R}$. Loosely speaking, the measure associates a random number
with any subset of $\mathbf{R}{\times} \mathbf{R}$. Whenever two
subsets are disjoint, the associated measures are independent, and the
measure of a disjoint union of sets almost certainly equals the sum of
the measures of the individual sets.  For a mathematically more
rigorous definition of \rm{i.s.r.m.'s} and their theory of
integration, see Refs.~\cite{SCH03b,KAL89,KWA92}.

Independently scattered random measures provide a natural basis for
describing uncorrelated noise processes in space and time. A special
class of \rm{i.s.r.m.'s} is that of homogeneous L\'evy bases, where
the distribution of the measure of each set is infinitely divisible
and does not depend on the location of the subset. In this case, it is
easy to handle integrals with respect to the L\'evy basis using the
well-known L\'evy-Khintchine and L\'evy-Ito representations for L\'evy
processes. Here, we state the result and point to \cite{SCH03b} for
greater detail and rigour.

Let $Z$ be a homogeneous L\'evy basis on
$\mathbf{R}{\times}\mathbf{R}$, i.e.\ $Z(S)$ is infinitely divisible
for any $S\subset\mathbf{R}{\times}\mathbf{R}$. Then we have the
fundamental relation
\begin{equation}
\label{fund}
\left \langle \exp \left \{ \int_{S}h(a) Z(\mathrm{d}a)\right \}
\right\rangle=
\exp \left \{ \int_{S} \mathrm{K}[h(a)]\mathrm{d}a \right \}
\,,
\end{equation} 
where $\langle \cdots \rangle$ denotes the expectation, $h$ is any
integrable deterministic function, and $\mathrm{K}$ denotes the
cumulant function of $Z(\mathrm{d}a)$, defined by
\begin{equation}
\ln \left \langle \exp \left \{ \xi Z(\mathrm{d}a) \right \}
\right \rangle=
\mathrm{K}[\xi]\,\mathrm{d}a.
\end{equation}
The usefulness of (\ref{fund}) is obvious: it permits explicit
calculation of the correlation function of the integrated and
$h$-weighted noise field $Z(\mathrm{d}a)$ once the cumulant function
$\mathrm{K}$ of $h$ is known.

\subsection{General model ansatz}     

Based on relation (\ref{fund}), we construct a spatio-temporal process
that is causal and continuous\footnote{In this context, continuity
  refers to the definition of observable $\epsilon(x,t)$ for a
  continuous range of points $(x,t)$.} by defining the observable
field $\epsilon(x,t)$ as
\begin{equation}
\label{def}
\epsilon(x,t)=\exp \left \{ \int_{S(x,t)}h(x,t;x',t')
Z(\mathrm{d}x'{\times} \mathrm{d}t')\right \}.
\end{equation}
This is clearly a multiplicative process of independent factors $\exp
\{h(x,t;x',t')\; Z(\mathrm{d}x'{\times} \mathrm{d}t')\}$ made up of a
specifiable \textit{weight function} $h$ and a homogeneous L\'evy
basis $Z$ over $\mathbf{R}{\times}\mathbf{R}$. Contributions to field
amplitude $\epsilon(x,t)$ lie within the influence domain $S(x,t)$,
called the associated ambit set.  To guarantee causality, we demand
that $S$ be nonzero only for times preceding the observation time $t$,
i.e.\ $S(x,t)\subset \mathbf{R}{\times}[-\infty,t]$ (see Figure 1).

Ansatz (\ref{def}) reduces to the model of Ref.~\cite{SCHMITT03} when
focusing on one purely temporal dimension, setting
$S(t)=[t+1-\lambda,t]$, $\lambda>1$, $h(t;t')=(t-t')^{-1/2}$ and
defining $Z$ to be Brownian motion. Similarly, a non-causal and again
purely temporal version of the general model (\ref{def}) with a
conical ambit set leads\footnote{
  This connection can be established by replacing the spatial
  coordinate $x$ with a scale label and omitting the causality
  condition.}
to the scale-dependent measures used in \cite{MUZ02}.

As shown in the Appendix and discussed in Section \ref{sec:linkmult},
our approach also includes the case of a multifractal measure that is
constructed without a limit-argument.  Moreover, it allows for
multifractality in space and time simultaneously. This multifractal
case (with the additional assumption of a stable L\'evy basis)
corresponds to the log-stable process described in \cite{SCH03a},
where the ambit set is constructed from an analogy to a cascade
process.  In Section \ref{sec:constr}, we derive the same result from
an alternative approach.

Some other applications of (\ref{def}) are discussed in \cite{SCH03b}.

The generality of the model (\ref{def}) is based on the possibility of
choosing the constituents of the process $\epsilon(x,t)$
independently.  The available degrees of freedom are the weight
function $h$, an arbitrary infinitely divisible distribution for the
L\'evy basis $Z$ (including Brownian motion, stable processes,
self-decomposable processes etc.) and the shape of the ambit set $S$.
As all of these quantities can be chosen to fit the purpose and
application in mind, our ansatz permits sensitive and flexible
modeling of the correlation structure of $\epsilon(x,t)$.  Despite its
generality, the model is tractable enough to yield explicit
expressions for arbitrary $n$-point correlations in closed form.

\subsection{n-point correlations}     

The definition of the process $\epsilon(x,t)$ allows for an explicit
calculation of arbitrary spatio-temporal $n$-point correlations,
defined as
\begin{equation}
\label{npoint}
c_{n}(x_{1},t_{1};\ldots ;x_{n},t_{n}) \equiv 
\left \langle \epsilon(x_{1},t_{1}) \cdot \ldots \cdot 
\epsilon(x_{n},t_{n})\right \rangle
\,,
\end{equation}
which give a complete characterisation of the correlation structure of
$\epsilon(x,t)$.

Using the definition (\ref{def}) and the fundamental relation
(\ref{fund}) we rewrite
\begin{eqnarray}
\label{cn}
&& c_{n}(x_{1},t_{1};\ldots ;x_{n},t_{n})= \nonumber \\
& = &  
\left \langle \exp \left \{ \sum_{i=1}^{n}
\int_{S(x_{i},t_{i})}h(x_{i},t_{i};x',t')
Z(\mathrm{d}x' {\times} \mathrm{d}t')\right \}\right \rangle
 \nonumber \\ \nonumber \\ & = & 
\left \langle \exp \left \{ \int_{\mathbf{R}{\times} \mathbf{R}}
\left(\sum_{i=1}^{n} \mathrm{I}_{S(x_{i},t_{i})} 
  h(x_{i},t_{i};x',t')\right) 
Z(\mathrm{d}x' {\times} \mathrm{d}t')\right \}\right \rangle \nonumber
\\  
\nonumber \\ & = &  
\exp \left \{ \int_{\mathbf{R}{\times} \mathbf{R}}
\mathrm{K}\left[\left(\sum_{i=1}^{n} \mathrm{I}_{S(x_{i},t_{i})} 
h(x_{i},t_{i};x',t')\right)\right] \mathrm{d}x'\mathrm{d}t'\right \},
\end{eqnarray}
where we made use of the index-function
\begin{equation}
  \mathrm{I}_{A}(x,t)=\left \{\begin{array}{ll}
      1 & \mathrm{when\ } (x,t)\in A\\ & \\
      0 & \mathrm{otherwise} \end{array}\right.
\end{equation}
for sets $A\subset \mathbf{R}{\times} \mathbf{R}$. The last step in
(\ref{cn}) follows from the fundamental equation (\ref{fund}).

To illustrate (\ref{cn}), we consider in more detail the cases $n=2$
and $n=3$ with the abbreviation $S_{i}=S(x_{i},t_{i})$. For $n=2$, it
follows that
\begin{eqnarray}
\label{n2}
\left \langle \epsilon(x_{1},t_{1}) \epsilon(x_{2},t_{2})\right \rangle 
& = & 
\exp \left \{ \int_{S_{1}\backslash S_{2}} 
\mathrm{K} \left[ h(x_{1},t_{1};x,t)\right] 
\mathrm{d}x\, \mathrm{d}t \right\}
\nonumber \\ 
& & \nonumber\\ 
&\times&
\exp \left \{ \int_{S_{2}\backslash S_{1}} 
\mathrm{K} \left[ h(x_{2},t_{2};x,t)\right] 
\mathrm{d}x\, \mathrm{d}t \right\} 
\nonumber \\ 
& & \nonumber\\ 
&\times&
\exp \left \{ \int_{S_{1}\cap S_{2}} 
\mathrm{K} \left[ h(x_{1},t_{1};x,t)+  h(x_{2},t_{2};x,t)\right] 
\mathrm{d}x\, \mathrm{d}t \right\}.
\end{eqnarray}
As illustrated in Figure 2.c, the first and second factor are
contributions from the non-overlapping parts of the ambit sets, while
the third stems from the overlap of $S(x_{1},t_{1})$ and
$S(x_{2},t_{2})$ (the shaded area). The latter factor describes the
correlation of the field amplitude $\epsilon$ at different
spatio-temporal locations; for locations where the overlap
$S(x_{1},t_{1})\cap S(x_{2},t_{2})$ vanishes, we get uncorrelated
field amplitudes $\langle \epsilon(x_{1},t_{1}) \epsilon(x_{2},t_{2})
\rangle- \langle \epsilon(x_{1},t_{1})\rangle\langle
\epsilon(x_{2},t_{2})\rangle=0$.  Thus, the extension and shape of
$S(x,t)$ characterises the range of correlations, while the overlap of
two ambits, the weight-function $h$ and the cumulant function
$\mathrm{K}$ influences the correlation strength.

In third order we get a similar result. The combinatorics of the
overlap of ambit sets for the observation points $(x_1, t_1)$, $(x_2,
t_2)$ and $(x_3, t_3)$ yields seven disjoint domains as follows: the
three domains $S(x_{1},t_{1})\backslash [S(x_{2},t_{2})\cup
S(x_{3},t_{3})]$, $S(x_{2},t_{2})\backslash [S(x_{1},t_{1})\cup
S(x_{3},t_{3})]$ and $S(x_{3},t_{3})\backslash [S(x_{1},t_{1})\cup
S(x_{2},t_{2})]$ give uncorrelated contributions associated solely
with one field amplitude; for instance, $S(x_{1},t_{1})\backslash
[S(x_{2},t_{2})\cup S(x_{3},t_{3})]$ is the contribution to
$\epsilon(x_{1},t_{1})$ that is independent of $\epsilon(x_{2},t_{2})$
and $\epsilon(x_{3},t_{3})$. A second set of three domains
$[S(x_{1},t_{1})\cap S(x_{2},t_{2})]\backslash S(x_{3},t_{3})$,
$[S(x_{1},t_{1})\cap S(x_{3},t_{3})]\backslash S(x_{2},t_{2})$ and
$[S(x_{2},t_{2})\cap S(x_{3},t_{3})]\backslash S(x_{1},t_{1})$
constitute the contributions to the correlation of two field
amplitudes but without that of the third field amplitude.  Finally,
$S(x_{1},t_{1})\cap S(x_{2},t_{2})\cap S(x_{3},t_{3})$ is the overlap
of all three ambit sets that describes the common correlation of all
three field amplitudes.

Using the simplified notation $\mathrm{K}_{i_{1},i_{2}.\ldots,i_{j}}
\equiv \mathrm{K}[h(x_{i_{1}},t_{i_{1}};x,t) +
h(x_{i_{2}},t_{i_{2}};x,t) + \cdots + h(x_{i_{j}},t_{i_{j}};x,t)]$,
the result in third order hence reads
\setlength{\jot}{12pt}
\begin{eqnarray}
\label{n3}
\lefteqn{\left \langle \epsilon(x_{1},t_{1}) \epsilon(x_{2},t_{2}) 
\epsilon(x_{3},t_{3})\right \rangle =}
\nonumber \\
&=&
\exp \left \{ \int_{S_{1}\backslash (S_{2}\cup S_{3}) } 
\!\!\!\!\!\mathrm{K}_{1}\, \mathrm{d}x\, \mathrm{d}t\right\} 
\exp \left \{ \int_{S_{2}\backslash (S_{1}\cup S_{3}) } 
\!\!\!\!\!\mathrm{K}_{2}\, \mathrm{d}x\, \mathrm{d}t\right\} 
\exp \left \{ \int_{S_{3}\backslash (S_{1}\cup S_{2}) } 
\!\!\!\!\!\mathrm{K}_{3}\, \mathrm{d}x\, \mathrm{d}t\right\} 
\nonumber\\ 
&\times&
\exp \left \{ \int_{(S_{1}\cap S_{2})\backslash S_{3} } 
\!\!\!\!\!\mathrm{K}_{1,2}\, \mathrm{d}x\, \mathrm{d}t\right\} 
\exp \left \{ \int_{(S_{1}\cap S_{3})\backslash S_{2} }
\!\!\!\!\!\mathrm{K}_{1,3}\, \mathrm{d}x\, \mathrm{d}t\right\} 
\exp \left \{ \int_{(S_{2}\cap S_{3})\backslash S_{1} }
\!\!\!\!\!\mathrm{K}_{2,3}\, \mathrm{d}x\, \mathrm{d}t\right\} 
\nonumber\\ 
&\times&
\exp \left \{ \int_{S_{1}\cap S_{2}\cap S_{3} }\!\!\! \!\!
\mathrm{K}_{1,2,3}\, \mathrm{d}x\, \mathrm{d}t\right\}\,.
\end{eqnarray}
It is clear from the above examples that the correlation structure
corresponds directly to an intuitive geometrical picture in which the
design and the overlap of the ambit sets $S$ determine the
correlation structure.

Conversely, one can use some given correlation structure $c_n$ as the
starting point for designing a suitable shape of the ambit set and
weight-function $h$ to fit these requirements, opening up a wide range
of applications. As outlined in the next section, multiscaling appears
as a specific example, while Ref.~\cite{SCH03b} provides further
insight into the kind of processes that can be modeled and explores
the potential of the additive counterpart defined as $\ln
\epsilon(x,t)$).

\section{Multiscaling model specifications}     
\label{sec:multiscal}

In this Section, the concept of multiscaling and multifractality is
examined in the context of the general model approach presented above.
An explicit expression for the ambit set $S$ is derived from scaling
two-point correlations, and fusion rules \cite{PRO96} expressing
$n$-point correlations solely in terms of scaling relations are
formulated.  Finally, the link to standard multifractality is
established.

\subsection{General remarks and assumptions}     

In order to keep the mathematics as transparent as possible, we will
use some simplifying assumptions about the structure of the process
$\epsilon(x,t)$. Our goal is the construction of a stationary and
translationally invariant process with scaling two-point correlations.
For the simplest way to achieve stationarity and translational
invariance, we assume $h\equiv 1$ and take the form of the ambit set
$S(x,t)$ to be independent of the location $(x,t)$, so that
\begin{equation}
\label{shape}
S(x,t)=(x,t)+S_{0}\,,
\end{equation}
where the shape of $S_{0}$ is independent of $(x,t)$. (Note that
$h\equiv 1$ is not a prerequisite for stationarity and translational
invariance. It would be sufficient to require $h(x',t';x,t)\equiv
h(x',t')$, but we can do without this additional degree of freedom for
the special case of scaling relations for two-point correlations.)

Figure 1 illustrates the various features of $S(x,t)$, which we now
discuss. At the origin $(0,0)$, it is specified mathematically by
\begin{equation}
\label{dmb}
S_{0}=\left\{ (x,t) \in \mathbf{R}{\times} \mathbf{R} 
: -T \le t \le 0, -g(t+T)\le x \le g(t+T)\right\}.
\end{equation}
This definition contains a finite decorrelation time $T$, ensuring
that no correlation survives for temporal separations $\Delta t$
larger than $T$, e.g.\ $\langle \epsilon(x,t)\epsilon(x,t+\Delta
t)\rangle- \langle \epsilon(x,t)\rangle \langle \epsilon(x,t+\Delta
t)\rangle=0$ for all $\Delta t\ge T$.

Spatially, the ambit $S_0$ is limited by a function $g(t)$, whose
monotonicity ensures that the spatial extension of the causality
domain increases monotonically for past times. The nonconstancy of $g$
implies a time-dependent spatial decorrelation length $l(\Delta t)$,
since, when two observations are separated by a space-time distance
$(\Delta x,\Delta t)$ (as illustrated in Figure 2.c), the two-point
correlation $\langle \epsilon(x,t)\epsilon(x+\Delta x,t+\Delta
t)\rangle- \langle \epsilon(x,t)\rangle \langle \epsilon(x+\Delta
x,t+\Delta t)\rangle$ vanishes for all $\Delta x \ge l(\Delta
t)=g(\Delta t)+g(0)$. The spatial decorrelation length $l(\Delta t)$
decreases monotonically with $\Delta t$, and its maximum
$l(0)=2g(0)\equiv L$ defines the decorrelation length $L$.  This is a
physically desirable property.

Finally, we impose a locality condition $g(T)=0$, i.e.\ the ambit set
$S_{0}$ is attached to $(x,t)$ in an unequivocal way.

The procedure followed in the next section starts from the assumption
that spatial and temporal two-point correlations scale, and constructs
the model according to this requirement. The basic relation we use in
the translationally invariant and stationary case under the
assumptions (\ref{shape}) and $h\equiv 1$ is
\begin{eqnarray}
\label{tp1}
\lefteqn{\langle \epsilon(x,t)\epsilon(x+\Delta x, t+\Delta t)\rangle }
\nonumber\\
&=&
\exp \left \{\int_{S_{1}\backslash S_{2}}\mathrm{K}[1]
\mathrm{d}x\,\mathrm{d}t\right\} \exp \left \{\int_{S_{2}\backslash S_{1}}
\mathrm{K}[1]\mathrm{d}x\,\mathrm{d}t\right\} 
\exp \left \{\int_{S_{1}\cap S_{2}}
\mathrm{K}[2]\mathrm{d}x\,\mathrm{d}t\right\} 
\nonumber \\ 
&=& \exp \left \{\int_{S_{1}}
\mathrm{K}[1]\mathrm{d}x\,\mathrm{d}t\right\} 
\exp \left \{\int_{S_{2}}
\mathrm{K}[1]\mathrm{d}x\,\mathrm{d}t\right\}\exp 
\left \{\int_{S_{1}\cap S_{2}}(\mathrm{K}[2]-
2\mathrm{K}[1])\mathrm{d}x\,\mathrm{d}t\right\}
\nonumber \\ 
&=& \langle \epsilon\rangle ^{2}
\exp \biggr\{ V(\Delta x, \Delta t)
(\mathrm{K}[2]-2\mathrm{K}[1])\biggr\} \,,
\end{eqnarray}
where we have used (\ref{n2}) with $h\equiv 1$ and the abbreviations
$S_{1}=S(x,t)$, $S_{2}=S(x+\Delta x, t+\Delta t)$ and
\begin{equation}
\label{vol}
V(\Delta x, \Delta t)=
\mathrm{Vol}(S(x,t)\cap S(x+\Delta x, t+\Delta t))
\end{equation}
for the Euclidean volume of the overlap of the ambit sets.  Due to
translational invariance and stationarity, we have $\langle
\epsilon(x,t) \rangle = \langle \epsilon(x+\Delta x, t+\Delta t)
\rangle = \langle \epsilon \rangle$.

Assuming that $\mathrm{K}[2]> 2\mathrm{K}[1]$ (note that, by the
strict convexity of log-Laplace transforms, we always have
$\mathrm{K}[2]-2\mathrm{K}[1] \ge 0$), Eq.~(\ref{tp1}) can be solved
for $V$,
\begin{equation}
\label{tp2}
V(\Delta x,\Delta t)=
\frac{1}{\mathrm{K}[2]-2\mathrm{K}[1]}
\;\;
{\ln \left( {\displaystyle \frac{\langle \epsilon(x,t)
\epsilon(x+\Delta x, t+\Delta t)\rangle}
{\langle \epsilon \rangle ^{2}}}\right)}.
\end{equation}
This relation establishes a simple geometrical way to design a model
with prescribed two-point correlations: one has only to choose ambit
sets $S(x,t)$ in a way that the volume of the overlap fulfils
(\ref{tp2}).  This will be done in the next Section for the case of
scaling two-point correlations (see \cite{SCH03b} for more examples
other than scaling relations).

\subsection{Construction of the ambit set  
via scaling two-point correlations}
\label{sec:constr}     

Implementing the general framework (\ref{def}) together with the above
assumptions and procedure, we start out by demanding power-law scaling
for the lowest-order spatial and temporal correlations,
\setlength{\jot}{3pt}
\begin{eqnarray}     
\label{eq:drei11}    
\left\langle \epsilon(x,t) \epsilon(x+\Delta x,t) \right\rangle 
= c_{x} (\Delta x)^{-\tau(2)}, \;\;\; \Delta x 
\in [l_{scal},L_{scal}] \subset[0,L],
\\ &&\nonumber\\
\label{eq:drei12}
\left\langle \epsilon(x,t) \epsilon(x,t+\Delta t) \right\rangle 
= c_{t} (\Delta t)^{-\tau(2)}, \;\;\; \Delta t 
\in [t_{scal},T_{scal}]\subset[0,T],
\end{eqnarray}     
with $c_{x}$ and $c_{t}$ constants. Note that the scaling exponents
$\tau(2)$ appearing in (\ref{eq:drei11}) and (\ref{eq:drei12}) are
taken to be identical; differing spatial and temporal scaling
exponents, as used previously e.g.\ in \cite{SCH02}, are easily
accommodated within our model, but do not satisfy the simpler
relations (\ref{simple}) given below.

Following the recipe sketched in (\ref{tp2}), we get, using
stationarity, for the temporal two-point correlation (\ref{eq:drei12})
the expression (see Figures 1 and 2.a)
\begin{eqnarray}
\label{d2}     
V(0,\Delta t)
&=&
\int_{\Delta t}^{T}2g(t)\mathrm{d}t \
= \int_{\Delta t}^{T-T_{scal}}2g(t)\mathrm{d}t \
+ \int_{T-T_{scal}}^{T}2g(t)\mathrm{d}t
\nonumber\\
&=&
\frac{\ln c_{t}-\ln (\langle \epsilon \rangle ^{2})}
     {\mathrm{K}[2]-2\mathrm{K}[1]}-\frac{\tau(2)\ln \Delta t}
{\mathrm{K}[2]-2\mathrm{K}[1]}
\end{eqnarray}     
for $\Delta t \in[t_{scal},T_{scal}]$, and after differentiation of
both sides with respect to $\Delta t$, we obtain the expression
\begin{equation}
\label{g}
g(t)=\frac{\tau(2)}{2(\mathrm{K}[2]-2\mathrm{K}[1])}\frac{1}{t},\;\;
\;\;\;t \in[t_{scal},T_{scal}]
\end{equation}
for the function $g(t)$ bounding the ambit set $S(x,t)$ within the
temporal scaling regime $[t_{scal},T_{scal}]$. The singularity of
$g(t)$ for $t\to 0$ and the locality condition $g(T)=0$
retrospectively justify the introduction of the cutoffs $t_{scal}$ and
$T_{scal}$ for the temporal scaling regime. We could also have started
from the spatial scaling relation (\ref{eq:drei11}) to obtain exactly
the same functional form for $g(t)$ with
\begin{equation}
\label{simple}
g(T_{scal})=\frac{l_{scal}}{2}, \;\;\; g(t_{scal})=\frac{L_{scal}}{2}.
\end{equation}
Thus the set of scaling relations (\ref{eq:drei11}) and
(\ref{eq:drei12}) are compatible under the assumption of a constant
weight-function $h\equiv 1$, i.e.\ there exists a solution for $g(t)$
that satisfies (\ref{eq:drei11}) and (\ref{eq:drei12}) simultaneously.
This sheds some light on the property of the weight-function $h$ to
select compatible temporal and spatial two-point correlations: scaling
relations are among the simplest functional forms and allow $h\equiv
1$, while for more advanced studies, such as deviations from scaling
for $\Delta t \notin [t_{scal},T_{scal}]$ and $\Delta x \notin
[l_{scal},L_{scal}]$, other weight-functions $h$ might be in order.
For a brief account of this topic we refer the reader to
\cite{SCH03b}.

To complete the specification of $g(t)$, functional forms in the time
intervals $0\leq t\leq t_{scal}$ and $T_{scal}\leq t\leq T$ are needed
in principle. For this analytical treatise, however, it is not
necessary to specify the functional form of $g(t)$ explicitly for
these two time intervals, since we neglect the constants of
proportionality $c_{x}$ and $c_{t}$ in (\ref{eq:drei11}) and
(\ref{eq:drei12}) in any case; examples addressing this issue can be
found in Ref.~\cite{SCH02}. The important point is that the validity
of the scaling relations (\ref{eq:drei11}) and (\ref{eq:drei12}) is
independent of a specific choice of $g(t)$ for $t\notin
[t_{scal},T_{scal}]$: Figure 1 and Figure 2.a show that, for purely
temporal separation, spatial scales in $S$ larger than $L_{scal}$ do
not contribute to the ambit overlap $S_1\cap S_2$ for $\Delta t
>t_{scal}$, while spatial scales in $S$ smaller than $l_{scal}$ are
completely part of the overlap for $\Delta t < T_{scal}$ and thus
contribute only a term constant in $\Delta t$.

Similar results hold for the purely spatial separation shown in Figure
2.b: regions of $S$ smaller than $l_{scal}$ do not contribute to the
overlap for $\Delta x >l_{scal}$. The contributions from the large
scales $>L_{scal}$ result in a constant, as can easily be seen from
\begin{eqnarray}
\label{svol}
V(\Delta x,0)
&=& \int_{0}^{g^{(-1)}(\Delta x/2)} 
\left(2g(t)-\Delta x\right)\mathrm{d}t
\nonumber\\
&=& \int_{0}^{t_{scal}}2g(t)\mathrm{d}t
\ +
\int_{t_{scal}}^{g^{(-1)}(\Delta x/2)} 2g(t)\mathrm{d}t
\ -
\frac{\tau(2)}{\mathrm{K}[2]-2\mathrm{K}[1]},
\end{eqnarray}
where $g^{(-1)}$ denotes the inverse of $g$.  Thus, a specific choice
of $g(t)$ for $t<t_{scal}$ only involves the constant $c_{x}$ and does
not influence the scaling behaviour (\ref{eq:drei11}) as such. The
only restriction is $V(0,0)<\infty$ (for finite expectations).

\subsection{Structure of higher-order correlations}     

In the previous Section, we specified the model starting from scaling
two-point correlations. It is now straightforward to derive scaling
relations for all higher order correlations of purely spatial and
temporal type. Section \ref{sec:linkmult} shows how these scaling
relations imply multifractality.

First we note that, since $h\equiv 1$, equation (\ref{cn}) translates
to
\begin{equation}
\label{star}
c_{n}(x_{1},t_{1};\ldots;x_{n},t_{n})=
\exp \left \{ \int_{\mathbf{R}{\times} \mathbf{R}}
\mathrm{K}\left [ \sum_{i=1}^{n} \mathrm{I}_{S(x_{i},t_{i})}(x,t)\right] 
\mathrm{d}x\, \mathrm{d}t\right \}.
\end{equation}
The argument of the cumulant function $\mathrm{K}$ in (\ref{star}) is
piecewise constant where $\sum_{i=1}^{n}
\mathrm{I}_{S(x_{i},t_{i})}(x,t)$ counts the number of field
amplitudes $\epsilon(x_{i},t_{i})$ that contribute to $(x,t)$ via
their ambit sets $S(x_{i},t_{i})$. This function vanishes outside of
$\bigcup_{i=1}^{n}S(x_{i},t_{i})$.

Focusing first on purely spatial two-point correlations of higher
order, we get, using (\ref{star}), the analog to (\ref{tp1})
\begin{eqnarray}
\label{eq:drei13}
\left \langle \epsilon(x,t)^{n_{1}}
\epsilon(x+\Delta x,t)^{n_{2}}\right \rangle 
& = & 
\left \langle \epsilon(x,t)^{n_{1}}\right \rangle 
\left \langle \epsilon(x+\Delta x,t)^{n_{2}}\right \rangle 
\nonumber\\
&\times&
\exp \left\{ V(\Delta x,0)\left(\mathrm{K}[n_{1}+n_{2}]-
\mathrm{K}[n_{1}]-\mathrm{K}[n_{2}]\right)\right\}.
\end{eqnarray}
Translational invariance and (\ref{g}), (\ref{svol}) imply scaling
relations for the higher-order two-point correlations
\begin{equation}
\label{scal}
\left \langle \epsilon(x,t)^{n_{1}}
\epsilon(x+\Delta x,t)^{n_{2}}\right \rangle
\propto (\Delta x)^{-\tau(n_{1},n_{2})},\;\;\; \Delta x 
\in [l_{scal},L_{scal}],
\end{equation}
where
\begin{equation}
\label{tau}
\tau(n_{1},n_{2})=\frac{\tau(2)}{\mathrm{K}[2]-2\mathrm{K}[1]}
\biggr(\mathrm{K}[n_{1}+n_{2}]
-\mathrm{K}[n_{1}]-\mathrm{K}[n_{2}]\biggr).
\end{equation}
(Again, the convexity of $\mathrm{K}$ implies
$\mathrm{K}[n_{1}+n_{2}]-\mathrm{K}[n_{1}]-\mathrm{K}[n_{2}] \ge 0$.)
The scaling range of (\ref{scal}) is identical to the scaling range of
(\ref{eq:drei11}) and does not depend on the order $(n_{1},n_{2})$.

An analogous procedure leads to scaling relations for the spatial
higher-order three-point correlations illustrated in Figure 3.  For
ordered points $x_1<x_2<x_3$ with relative distances assumed to be
within the spatial scaling range,
$|x_{i}-x_{j}|\in[l_{scal},L_{scal}]$, $(i,j=1,2,3)$, we find that
\setlength{\jot}{12pt}
\begin{eqnarray}     
\label{eq:drei24} 
&&
  \left\langle  
  \epsilon(x_1,t)^{n_1} 
  \epsilon(x_2,t)^{n_2} 
  \epsilon(x_3,t)^{n_3} 
  \right\rangle
\nonumber\\ 
&\propto&  \left(x_2-x_1\right)^{-\tau(n_1,n_2)}  
          \left(x_3-x_2\right)^{-\tau(n_2,n_3)} 
          \left(x_3-x_1\right)^{-\xi(n_1,n_2,n_3)}     
          \,, 
\end{eqnarray}  
with a modified exponent $\xi$ defined by 
\begin{equation}     
\label{eq:drei25}     
  \xi(n_1,n_2,n_3) 
    =  \tau(n_1+n_2,n_3) - \tau(n_2,n_3) 
       \; . 
\end{equation}     
The reason for the different forms of the exponents $\tau$ and $\xi$
lies, of course, in the different ambit set overlaps: as shown in
Figure 3, points $x_{1}$ and $x_{3}$ have only the one neighbour
$x_{2}$, while $x_{2}$ has two.

Equation (\ref{eq:drei24}) can be viewed as a generalised fusion rule
in the sense of \cite{PRO96}. It is easily generalised to $n$-point
correlations of arbitrary order because all overlapping ambit sets can
be written as a combination of overlaps $V(|x_{i}-x_{j}|,0)\propto \ln
|x_j-x_i|$, as long as $|x_i-x_j| \in[l_{scal},L_{scal}]$ for all
point pairs.  As shown by induction in \cite{SCH02}, the spatial
$n$-point correlation for ordered points $x_1 <x_2 < \ldots < x_n$ and
arbitrary order $(m_{1}, \ldots ,m_{n})$ satisfying $x_{i+1}-x_i \in
[l_{scal},L_{scal}]$ has the following structure:
\setlength{\jot}{12pt}
\begin{eqnarray} 
\label{spatio}    
&&  \left\langle  
  \epsilon(x_1,t)^{m_1} \cdots \epsilon(x_n,t)^{m_n} 
  \right\rangle 
\nonumber\\
&\propto&  \left( \prod_{i=1}^{n-1}     
          \left( x_{i+1}-x_i \right)^{-\tau(m_i,m_{i+1})}  
          \right)  
          \prod_{j=2}^{n-1} \prod_{l=j+1}^{n}  
          \left( x_l-x_{l-j} \right)^{-\xi(m_{l-j},\ldots ,m_l)},     
\end{eqnarray}
where
\begin{equation}     
\label{eq:drei27}     
  \xi(m_{l-j},\ldots,m_l)  
    =  \tau(m_{l-j}+\ldots+m_{l-1},m_l) 
    - \tau(m_{l-j+1}+\ldots+m_{l-1},m_l)     
    \,. 
\end{equation}     
The modified scaling-exponents $\xi(m_1,\ldots,m_j)$ correspond to
(\ref{eq:drei25}) for $j=3$ and arise from the nested structure of the
overlapping ambit sets. Physically, Eq.~(\ref{spatio}) implies that
spatial $n$-point correlations factorise into contributions arising at
the smallest scales $x_{i+1}-x_i$, at next-to-smallest scales
$x_{i+2}-x_i$,  and so on up to the largest scale, $x_n-x_1$.

To complete the discussion of $n$-point correlations, we state the
corresponding relation for temporal $n$-point correlations of
arbitrary order
\setlength{\jot}{12pt}
\begin{eqnarray}     
\label{eq:drei26}     
&&  \left\langle  
  \epsilon(x,t_1)^{m_1} \cdots \epsilon(x,t_n)^{m_n} 
  \right\rangle 
\nonumber\\
&\propto&  \left( \prod_{i=1}^{n-1}     
          \left( t_{i+1}-t_i \right)^{-\tau(m_i,m_{i+1})}  
          \right)  
          \prod_{j=2}^{n-1} \prod_{l=j+1}^{n}  
          \left( t_l-t_{l-j} \right)^{-\xi(m_{l-j},\ldots ,m_l)} 
          \; ,     
\end{eqnarray}     
for ordered times $t_1< \ldots <t_n$ and $|t_{i}-t_{j}|\in
[t_{scal},T_{scal}]$, $i,j=1,\ldots ,n$.  Finally it is to be noted
that relations (\ref{spatio}) and (\ref{eq:drei26}) only hold for
purely spatial and purely temporal $n$-point correlations
respectively. The general case of arbitrary $n$-point correlations
(\ref{star}) does not allow a similar description in terms of scaling
relations, since $V(\Delta x,\Delta t)$ includes mixed terms in
$\Delta x$ and $\Delta t$. For a complete discussion of general
space-time two-point correlations, we refer again to \cite{SCH02}.

\subsection{Link to classical multifractality}     
\label{sec:linkmult}

We complete the discussion of the multiscaling model with an
investigation of the relation between multiscaling (defined as scaling
of $n$-point correlations (\ref{spatio}) and (\ref{eq:drei26})) and
classical multifractality (defined as scaling of coarse-grained
moments).  In the Appendix, we prove that multiscaling implies
multifractality in the large scale limit.

The term multifractality in the classical sense refers to $n$-th order
moments of the field, coarse-grained at scale $l$ centered on
locations $\sigma$, displaying scaling behaviour with some non-linear
multifractal scaling exponent $\mu(n)>0$,
\begin{eqnarray}     
\label{LM1}     
M_{n}(\sigma,l)=\left \langle \left(\frac{1}{l}
\int_{\sigma-l/2}^{\sigma+l/2} \epsilon(\sigma') 
\mathrm{d}\sigma'\right)^{n} \right \rangle \propto l^{-\mu(n)}.
\end{eqnarray}
Note that this relation applies to stationary processes
$\epsilon(\sigma)$ since the right hand side of (\ref{LM1}) is
independent of the location $\sigma$. Differentiating this relation
twice with respect to $l$, it follows in second order, due to
stationarity, that the two-point correlations
\begin{eqnarray}     
\label{LM2}     
\left \langle \epsilon(\sigma+l)\epsilon(\sigma)\right \rangle 
\propto l^{-\mu(2)}     
\end{eqnarray}
scale with the same scaling exponent $\mu(2)$ as $M_2$. The inverse
need not be true, for the scaling relation (\ref{LM2}) becomes
singular for $l \rightarrow 0$, though at small scales deviations from
(\ref{LM2}) have to occur which in turn may destroy the relation
(\ref{LM1}) \cite{WOLF00,CLEVE03}.  However, (\ref{LM2}) indicates a
strong connection between scaling of $n$-point correlations and
multifractal scaling of order $n$.

The multiscaling model implies scaling relations for $n$-point
correlations, with deviations from pure scaling for scales smaller
than $l_{scal}$ for spatial correlations and scales smaller than
$t_{scal}$ for temporal ones.  Thus the question arises whether
multifractal exponents $\mu(n)$ are to be expected (see also
\cite{SCH02}). To answer this question, we assume the one-point
moments $\langle \epsilon(x,t)^{n}\rangle$ to be finite (i.e. we
restrict to L\'evy bases with $\mathrm{K}[n] <\infty$).

In the Appendix it is shown that the integral moments of temporal type
\begin{eqnarray} 
\label{tmom}    
M^{(t)}_{n}(t,l)=\left \langle \left(\frac{1}{l}\int_{t-l/2}^{t+l/2}     
\epsilon(x,t')\mathrm{d}t'\right)^{n} \right \rangle 
\propto l^{-\mu(n)}
\end{eqnarray}
asymptotically exhibit scaling behaviour for $t_{scal} \ll l$.
Moreover this is also true for the integral moments of spatial type
\begin{eqnarray} 
\label{smom}    
M^{(s)}_{n}(x,l)=\left \langle \left(\frac{1}{l}\int_{x-l/2}^{x+l/2}      
\epsilon(x',t)\mathrm{d}x'\right)^{n} \right \rangle 
\propto l^{-\mu(n)}
\end{eqnarray}
for $l_{scal} \ll l$ with the same multifractal scaling exponents
\begin{eqnarray}     
\mu(n)=\frac{\tau(2)}{\mathrm{K}[2]-
2\mathrm{K}[1]}(\mathrm{K}[n]-n\mathrm{K}[1]).
\end{eqnarray}

The crucial assumption that enters the proof of (\ref{tmom}) and
(\ref{smom}) is
\begin{equation}
\label{cond}
\tau(2)\frac{\mathrm{K}[n]-\mathrm{K}[n-1]-\mathrm{K}[1]}{\mathrm{K}[2]
-2\mathrm{K}[1]}
=\mu(n)-\mu(n-1)<1.
\end{equation}
This assumption ensures that large scale correlations dominate the
moments of the coarse grained field. It is to note that (\ref{cond})
is a sufficient condition for multifractality in the large-scale
limit. The statistics of the energy dissipation in fully developed
turbulence is an important example of an observable where condition
(\ref{cond}) holds; see Ref.~\cite{sreeni}.

The identity of spatial and temporal multifractal scaling exponents
$\mu(n)$ is clearly a result of the identical scaling behaviour of
purely spatial and temporal $n$-point correlations. The scaling of
spatial and temporal integral moments is independent of the choice of
the boundary function $g(t)$ for $t\notin[t_{scal},T_{scal}]$ as long
as $V(0,0)<\infty$. Under these mild restrictions, we are able to
model a wide range of scaling exponents $\mu(n)$ by choosing a proper
cumulant function $\mathrm{K}$ via the L\'evy basis $Z$ that fulfils
the sufficient condition (\ref{cond}). Examples are $\mu(n)\propto
n^{\alpha}-n$ for a stable basis with index of stability $0<\alpha \le
2$, $\alpha \neq 1$ and
\begin{equation}
\mu(n)\propto (1-n)\sqrt{\alpha^{2}-\beta^{2}}+n\sqrt{\alpha^{2}-
(\beta+1)^{2}}-\sqrt{\alpha^{2}-(\beta+n)^{2}},
\end{equation}    
with $|\beta+n|\le \alpha$, for a normal-inverse-Gaussian distribution
$\mathrm{NIG}(\alpha,\beta,\delta,\nu)$ \cite{BAR78,BAR98a,BAR98b}.
Depending on the parameters that characterize the distributions, there
exists a critical order $n_{c}$ where (\ref{cond}) does not hold any
more. The $\mathrm{NIG}(\alpha,\beta,\delta,\nu)$ distribution is an
example of a L\'evy basis where multifractality (\ref{LM1}) is defined
only up to a finite order $n$, since $\mathrm{K}[n] <\infty$ only for
$|\beta+n|\le \alpha$; for larger $n$, the moments $\langle
\epsilon^{n}\rangle $ and $M_{n}$ do not exist.

\section{Conclusion}     
\label{sec:concl}

We have presented a general framework for modeling of spatio-temporal
processes that allows, even in its generality, an analytical treatment
of general spatio-temporal $n$-point correlations. This framework
consists of a homogeneous L\'evy basis, the concept of an ambit set as
an associated influence domain and a weight-function $h$.  These three
degrees of freedom can be chosen arbitrarily and independently, thus
encompassing a wide range of applications. In this respect, we
mentioned briefly related work \cite{SCHMITT03,MUZ02,SCH03a} and
showed them to be special cases of this framework. In a specific
illustration, we have shown that a stationary and translationally
invariant version of the general model can be used to construct a
multiscaling and multifractal causal spatio-temporal process starting
from scaling relations of two-point correlations.

Many applications immediately come to mind. The great flexibility and
tractability of the framework's mathematics might well find its way
into modeling of rainfields, cloud distributions and various growth
models, to name just a few examples of spatio-temporal processes.
Another field of application for the special case of the multiscaling
model is the description and modeling of the statistics of the
energy-dissipation in fully developed turbulence as a prototype of a
multifractal and multiscaling field. A first step in this direction
was undertaken in \cite{SCH03a}, where scaling two- and three-point
correlations (\ref{eq:drei12}) and (\ref{eq:drei26}) were shown to be
in excellent correspondence with data extracted from a turbulent shear
flow experiment.

\section*{Acknowledgements}
  
This work was supported in part by the EU training research network
DynStoch -- Statistical methods for Dynamical Stochastic Models.
O.E.B.-N.\ and J.S.\ acknowledge support from MaPhySto -- A Network in
Mathematical Physics and Stochastics, funded by the Danish National
Research Foundation and support from the Alexander von Humboldt
Foundation with a Feodor-Lynen-Fellowship. H.C.E. acknowledges support
from the South African National Research Foundation.

\section*{Appendix: 
Scaling relations for integral moments}

This appendix proves the classical multifractal property (\ref{LM1})
for the multiscaling model in the limit $t_{scal} \ll t \le T_{scal}$
and $l_{scal} \ll l \le L_{scal}$ under the assumption that
\begin{equation}
\label{condA}
\tau(2)\frac{\mathrm{K}[n]-\mathrm{K}[n-1]-\mathrm{K}[1]}{\mathrm{K}[2]
-2\mathrm{K}[1]}<1.
\end{equation}
The proof is carried out in detail only for the spatial case; the
temporal counter part of the above statement is straightforward.

With the abbreviation $d_{n}(l_{2},\ldots,l_{n})=\langle \epsilon(0,t)
\epsilon(l_{2},t)\cdots \epsilon(l_{n},t)\rangle$ , $0<l_{2}<\ldots
l_{n}$ for the spatial correlation function of order $n$ and using the
translational invariance of the correlation structure, the spatial
integral moments of order $n$ (\ref{smom}) are given by
\begin{eqnarray}
\label{m0}     
M_{n}^{(s)}(x,l)
=n! \;l^{-n} \int_{0}^{l}
\mathrm{d}l_{n}\int_{0}^{l_{n}}\mathrm{d}l_{n-1}\cdots     
\int_{0}^{l_{3}}
\mathrm{d}l_{2}\left(l-l_{n}\right)d_{n}\;(l_{2},\ldots , l_{n}).     
\end{eqnarray}
To calculate the involved overlaps of the influence domains, it must
be distinguished whether the spatial distances are smaller or larger
than $l_{scal}$.  In the limit $l_{scal} \ll l$, the dominant
contribution is
\setlength{\jot}{12pt}
\begin{eqnarray}     
\label{m1}     
&&
M_{n}^{(s)}(x,l)\approx \tilde{M}_{n}^{(s)}(l) 
\\
&=& 
n! \;l^{-n} \int_{(n-1)l_{scal}}^{l}dl_{n}
\int_{(n-2)l_{scal}}^{l_{n}-l_{scal}}
dl_{n-1}\cdots     
\int_{l_{scal}}^{l_{3}-l_{scal}}dl_{2}\;
\left(l-l_{n}\right)d_{n}(l_{2},\ldots , l_{n}).     
\nonumber
\end{eqnarray}
The proof of the multifractality of $M_{n}^{(s)}(x,l)$ is carried out
in two steps. The first part shows that $\tilde{M}_{n}^{(s)}(l)\propto
l^{-\mu(n)}$ in the large scale limit. In the second step, we show
that the approximation $M_{n}^{(s)}(x,l)\approx
\tilde{M}_{n}^{(s)}(l)$ holds for $l\gg l_{scal}$. We also provide a
rough estimate for the relative error $|M_{n}^{(s)}(x,l)-
\tilde{M}_{n}^{(s)}(l)|/ \tilde{M}_{n}^{(s)}(l)$.

The correlation function $d_{n}$ can be rewritten with the help of the
generalised fusion rules (\ref{spatio}) as
\begin{eqnarray}     
\label{m2}     
d_{n}(l_{2},\ldots,l_{n})\propto\prod_{k=2}^{n}
\prod_{j=1}^{k-1}\left(l_{k}-l_{k-j}\right)^{-\xi_{j+1}}     
\end{eqnarray}
where 
\begin{equation}
\xi_{j+1}=\xi(\underbrace{1,\ldots,1}_{j-\rm{times}})
\end{equation} 
and $\xi_{2}\equiv \tau(1,1)$. In the next step, we define
\begin{eqnarray}
\label{FN}
&&
F_{n}(l,l_{scal})
\\
&\equiv&
l^{-n} \int_{(n-1)l_{scal}}^{l}dl_{n}\int_{(n-2)l_{scal}}^{l_{n}-l_{scal}}
dl_{n-1}\cdots     
\int_{l_{scal}}^{l_{3}-l_{scal}}dl_{2}
\left(l-l_{n}\right) \prod_{k=2}^{n}
\prod_{j=1}^{k-1}\left(l_{k}-l_{k-j}\right)^{-\xi_{j+1}}.
\nonumber
\end{eqnarray}
Note that $\tilde{M}_{n}(l)\propto F_{n}(l,l_{scal})$ with a
constant of proportionality that is independent of $l$.

With the abbreviation 
\begin{equation}
\label{h1}
h(k)=-\sum_{j=1}^{k-1}\xi_{j+1},
\end{equation}
it follows from (\ref{eq:drei25} and (\ref{eq:drei27}) that
\begin{equation}
\label{h2}
\sum_{k=2}^{n}h(k)=-\mu(n)
\end{equation}
where
\begin{equation}
\label{h3}
\mu(n)=\tau(2)\frac{\mathrm{K}[n]-n\mathrm{K}[1]}
{\mathrm{K}[2]-2\mathrm{K}[1]}.
\end{equation}
Thus we get, using condition (\ref{condA})
\begin{equation}
\label{h4}
h(n)
=\mu(n-1)-\mu(n)
=-\tau(2)\frac{\mathrm{K}[n]-\mathrm{K}[n-1]-\mathrm{K}[1]}{\mathrm{K}[2]
-2\mathrm{K}[1]}>-1.
\end{equation}
It follows immediately that 
\begin{equation}
\label{pro1}
\tilde{M}_{n}\,(l)l^{\mu(n)}\propto F_{n}(1,l_{scal}/l).
\end{equation}
$F_{n}(1,l_{scal}/l)$ is positive and increasing with increasing $l$.
It is easy to show that $F_{n}(1,l_{scal}/l)$ is bounded. From
(\ref{m2}) and (\ref{h1}) it follows that $d_{n}<
\prod_{k=2}^{n}l_{k}^{h(k)}$ and therefore
\begin{equation}
\label{bound}
F_{n}(1,l/l_{scal})< \int_{l_{scal}/l}^{1}
\mathrm{d}l_{n}\int_{l_{scal}/l}^{1}\mathrm{d}l_{n-1}\ldots
\int_{l_{scal}/l}^{1} \mathrm{d}l_{2}\prod_{k=2}^{n}l_{k}^{h(k)}
\le \prod_{k=1}^{n}\frac{1}{1+h(k)}.
\end{equation}
The last step in (\ref{bound}) requires (\ref{condA}) to hold. Since
$F_{n}(1,l_{scal}/l)$ is increasing with $l$ and bounded, there exists
a constant $c$ with
\begin{equation}
\label{lim}
\lim_{l\rightarrow \infty}\tilde{M}_{n}(l)l^{\mu(n)}=c<\infty
\end{equation}
and therefore
\begin{equation}
\label{scalA}
\tilde{M}_{n}(l)\propto l^{-\mu(n)}
\end{equation}
in the large scale limit $l\gg l_{scal}$.

To complete the calculations, we provide a rough estimate of the
relative error between the exact relation (\ref{m0}) and its
approximation (\ref{m1}).  By going from (\ref{m0}) to (\ref{m1}) we
neglect all $n$-point correlations with one or more distances
$|l_{i}-l_{j}| <l_{scal}$. These are ${n \choose 1}$ integrals of the
form
\setlength{\jot}{0pt}
\begin{eqnarray}
\label{ex}
&& \\
n! l^{-n} \int_{(n-1)l_{scal}}^{l}
\mathrm{d}l_{n} \ldots \int_{il_{scal}}^{l_{i+2}-l_{scal}}
\mathrm{d}l_{i+1}\int_{l_{i+1}-l_{scal}}^{l_{i+1}}
\mathrm{d}l_{i}\int_{(i-2)l_{scal}}^{l_{i}-l_{scal}}
\mathrm{d}l_{i-1}\ldots 
\int_{l_{scal}}^{l_{3}-l_{scal}}
\mathrm{d}l_{2}(l-l_{n})d_{n}(l_{2},\ldots,l_{n}),
\nonumber
\end{eqnarray}
where one distance (chosen to be $l_{i+1}-l_{i}$ in (\ref{ex})) is
smaller $l_{scal}$ and ${n \choose 2}$ integrals where two distances
are simultaneously smaller $l_{scal}$ etc., and one integral where all
distances are smaller than $l_{scal}$. Each of these integrals have an
upper bound $l_{scal}^{k} l^{n-k} d_{n}(0,\ldots,0)$ (assumed to be
finite), where $k$ denotes the number of distances that are smaller
$l_{scal}$.  Thus we have
\setlength{\jot}{3pt}
\begin{eqnarray}
\left| M_{n}^{(s)}(x,l)- \tilde{M}_{n}^{(s)}(x,l)\right | 
& \le &  
n!l^{-n}\sum_{k=1}^{n} {n \choose k}l_{scal}^{k} 
l^{n-k}d_{n}(0,\ldots,0)
\nonumber\\
&=& n!l^{-n}d_{n}(0,\ldots,0)
\left\{\left(l_{scal}+l\right)^{n}-l^{n}\right\}.
\end{eqnarray}
The relative error
\begin{equation}
\label{re}
\frac{\left| M_{n}^{(s)}(x,l)- \tilde{M}_{n}^{(s)}(x,l)\right|}
{ \tilde{M}_{n}^{(s)}(x,l)}\le \textrm{const.}\times 
l^{\mu(n)-n}\left\{ \left(l_{scal}+l\right)^{n}-l^{n}\right \}
\end{equation} 
tends to zero for $l\rightarrow \infty$ and $n>\mu(n)$ (which is
always true, for $l^{n}M_{n}^{(s)}(x,l)$ is monotonically increasing
for positive and finite $n$-point correlations).  The results
(\ref{scalA}) and (\ref{re}) are independent of the choice of the
small scale statistics as long as they are finite.


\newpage     
\pagestyle{empty}     
\begin{figure}     
\begin{centering}     
\epsfig{file=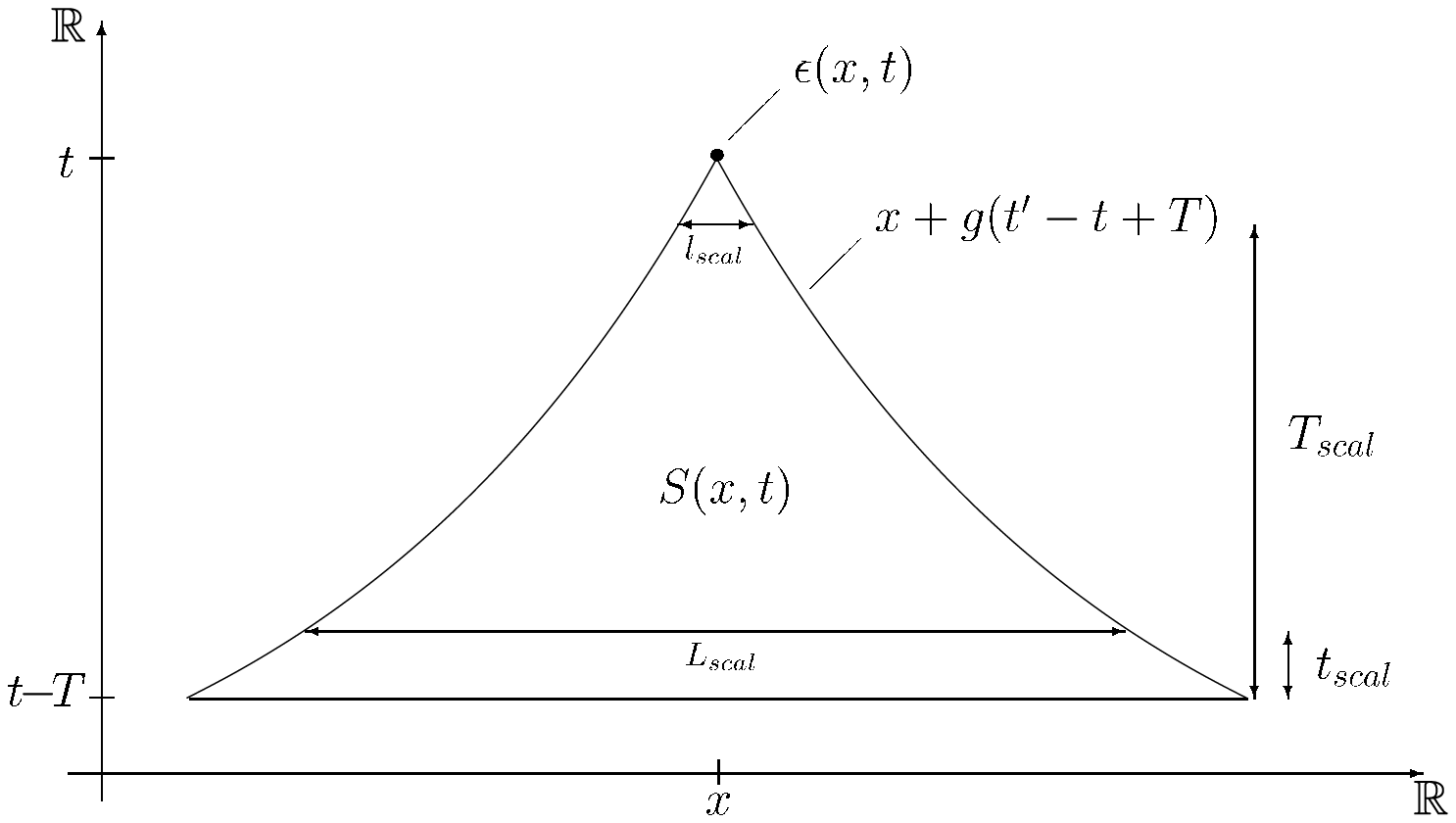,width=15cm}     
\caption{Illustration of the spatio-temporal ambit set $S(x,t)$
  associated with the field amplitude $\epsilon(x,t)$ and bounded by a
  monotone function $g(t'-t+T)$.  }
\end{centering}     
\end{figure}

\begin{figure}     
\begin{centering}     
\epsfig{file=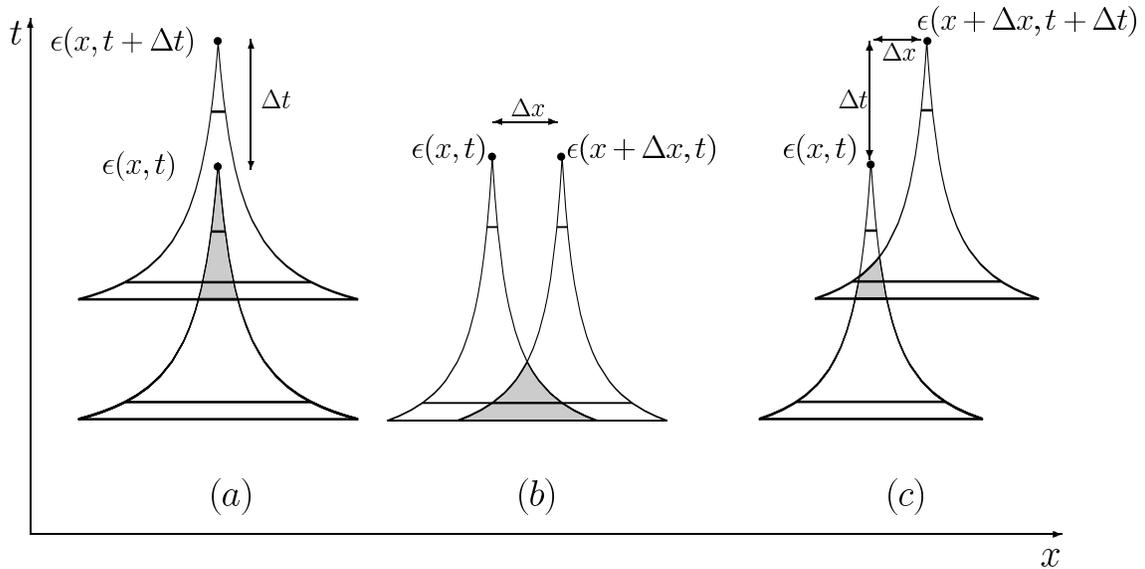,width=15cm}     
\caption{Spatio-temporal overlaps (shaded areas) of the ambit sets
  separated by a (a) temporal distance $\Delta t$, (b) spatial
  distance $\Delta x$ and (c) spatio-temporal distance $(\Delta
  x,\Delta t)$.  }
\end{centering}     
\end{figure}  
     
\begin{figure}     
\begin{centering}     
\epsfig{file=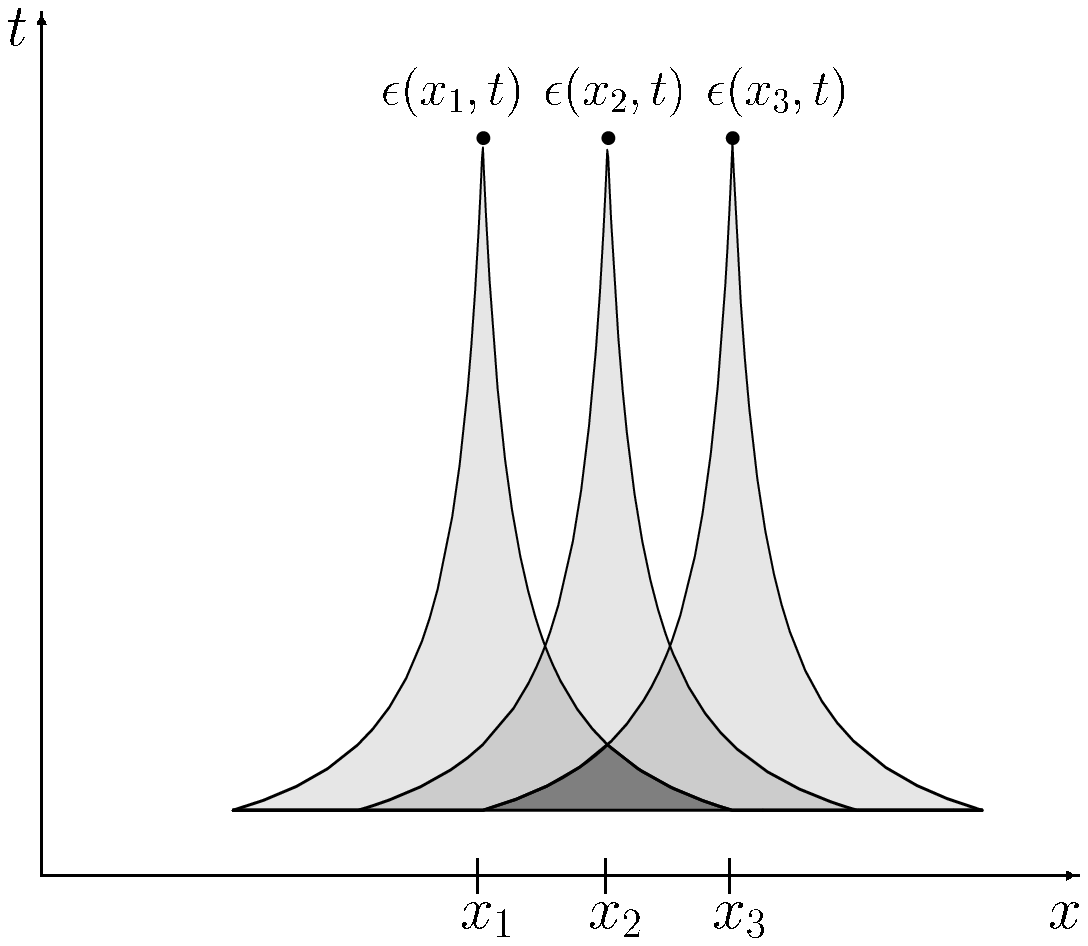,width=10cm}     
\caption{Illustration of the six disjoint contributions to the
  equal-time three-point correlation $\langle \varepsilon(x_1,t)
  \varepsilon(x_2,t) \varepsilon(x_3,t) \rangle$.  }
\end{centering}     
\end{figure}


\begin{thebibliography}{99}
\bibitem{FED88}
         Feder, J. (1988): \emph{Fractals}. 
         Plenum Press, New York.
\bibitem{MEN91}
         Meneveau, C. and Sreenivasan, K.R. (1991): 
         The multifractal nature of turbulent energy dissipation, 
         \textit{J.\ Fluid Mech}. \textbf{224}, 429-484.
\bibitem{FRI95}
         Frisch, U. (1995): 
         \emph{Turbulence. The legacy of A.N. Kolmogorov}. 
         Cambridge University Press.
\bibitem{PAR00}
         Park, K. and Willinger, W. (2000): 
         \emph{Self-similar network traffic and performance evaluation}. 
         John Wiley \& Sons, New York.
\bibitem{SCHER92}
          Lovejoy, S., Schertzer, D. and Watson, B. (1992): 
          Radiative Transfer and Multifractal Clouds: 
          theory and applications.
          \textit{I.R.S.} \textbf{92}, A. Arkin et al., Eds., 108-111.
\bibitem{SCHER85}
         Schertzer, D. and Lovejoy, S. (1985): 
         Generalized scale invariance in turbulent phenomena.
         \textit{Physico-Chemical Hydrodynamics Journal}  
         \textbf{6}, 623-635.
\bibitem{BRU98}     
        Br\'u, A., Pastor, J.M., Fernaud, I., Br\'u, I., Melle, S. 
        and Berenguer, C. (1998): 
        Super-rough dynamics on tumor growth.     
        \textit{Phys. Rev. Lett.} \textbf{81}, 4008-4011.
\bibitem{MUZ00}
         Muzy, J.F., Delour, J. and Bacry, E. (2000): 
         Modelling fluctuations of financial time series: 
         from cascade process to stochastic volatility model.
         \textit{Eur. Phys. J. B} \textbf{17}, 537-548.
\bibitem{CAL02}
        Calvet, L. and Fisher, A. (2002):
        Multifractality in Asset Returns: Theory and Evidence.
        \textit{The Review of Economics and Statistics}  
        \textbf{84}, 381-406.
\bibitem{BAR01}
         Barndorff-Nielsen, O.E. and Prause, K. (2001): 
         Apparent scaling.
         \textit{Finance Stochast}. \textbf{5}, 103-113.
\bibitem{SCHMITT03}     
        Schmitt, F.G. (2003): 
        A causal multifractal stochastic equation and its 
        statistical properties. Preprint      
        \textit{cond-mat}/0305655.
\bibitem{MUZ02}     
        Muzy, J.F. and Bacry, E. (2002):
        Multifractal stationary random measures and multifractal 
        random walks with log infinitely divisible scaling laws.     
        \textit{Phys. Rev. E} \textbf{66}, 056121.
\bibitem{SCH03a}
        Schmiegel, J., Cleve, J., Eggers, H.C., Pearson, B.R. 
        and Greiner, M. (2003): 
        Stochastic energy-cascade model for 1+1 dimensional 
        fully developed turbulence. 
        \textit{Phys. Lett. A} \textbf{320}, 247-253.
\bibitem{SCH03b}
        Barndorff-Nielsen O.E. and Schmiegel, J. (2003): 
        Levy-based Tempo-Spatial Modelling; 
        with Applications to Turbulence.
        Uspekhi Mat. Nauk \textbf{159} 63.
\bibitem{SCH02}
         Schmiegel, J. (2002): Ein dynamischer Prozess f\"ur die
         statistische Beschreibung der Energiedissipation in der 
         vollentwickelten Turbulenz. Dissertation TU Dresden, Germany.
\bibitem{KAL89} 
         Kallenberg, O. (1989): \emph{Random Measures}. (4th Ed.) Berlin:
         Akademie Verlag.
\bibitem{KWA92} 
         Kwapien, S. and Woyczynski, W.A. (1992): 
         \textit{Random Series and Stochastic Integrals: 
           Single and Multiple}, Basel: Birkh\"{a}user.
\bibitem{PRO96}     
         L'vov, V. and Procaccia, I. (1996): 
         Fusion rules in turbulent systems with flux equilibrium.     
         \textit{Phys. Rev. Lett.} \textbf{76}, 2898-2901.
\bibitem{WOLF00}     
         Wolf, M., Schmiegel, J. and Greiner, M. (2000):
         Artificiality of multifractal phase transitions.
         \textit{Phys. Lett. A} \textbf{266}, 276-281.
\bibitem{CLEVE03}     
         Cleve, J., Greiner, M. and Sreenivasan, K.R. (2003):
         On the surrogacy of the energy dissipation field in 
         fully developed turbulence.
         \textit{Europhys. Lett} \textbf{61}, 756-761.
\bibitem{sreeni} 
Sreenivasan, K.R. and Antonia, R.A. (1997) The
phenomenology of small-scale turbulence. \textit{Ann. Rev. Fluid Mech}. 
\textbf{29}, 435-472.
\bibitem{BAR78}     
         Barndorff-Nielsen, O.E. (1978): 
         Hyperbolic distributions and distributions on hyperbolae.
         \textit{Scand. J. Statist.} \textbf{5}, 151-157.
\bibitem{BAR98a}     
         Barndorff-Nielsen, O.E. (1998a): 
         Processes of normal inverse Gaussian type.
         \textit{Finance and Stochastics} \textbf{2}, 41-68.
\bibitem{BAR98b}     
         Barndorff-Nielsen, O.E. (1998a): 
         Probability and Statistics; selfdecomposability, 
         finance and turbulence. In L. Accardi and C.C. Heyde (Eds.): 
         Proceedings of the Conference 
         ``\textit{Probability towards 2000}'', 
         held at Columbia University, New York, 2-6 October 1995. 
         Berlin: Springer-Verlag. 47-57.

\end{thebibliography}
\end{document}